\documentclass[aps,notitlepage,twocolumn,nofootinbib,floatfix,pra,10pt]{revtex4-2}

\usepackage{amsmath,amsfonts,amssymb,amsthm,bbm,graphicx,times,mathtools,physics}
\usepackage{hyperref}
\usepackage{enumitem}

\hypersetup{colorlinks,linkcolor=blue,citecolor=red,urlcolor=blue}

\usepackage[ruled,longend]{algorithm2e}

\usepackage{xcolor}
\usepackage{todonotes}

\begin{document}

\title{To Scale Up or To Scale Out: Evaluating Space-Time Costs of Compiled Logical Circuits on Modular Superconducting Quantum Processors}

\author{Nikiforos Paraskevopoulos$^{1}$}
\author{S\'{e}bastian de Bone$^{1}$}
\author{Mick Christophersen$^{2}$}
\author{Simon Storz$^{3,4}$}
\author{A. Mert Bozkurt$^{2}$}
\email{mert.bozkurt@quantware.com}
\author{Arno Bargerbos$^{2}$}
\email{arno@quantware.com}
\author{Sebastian Feld$^{6,7}$}

\affiliation{$^{1}$Dark Qore, Poortweg 4, 2612 PA Delft, The Netherlands}
\affiliation{$^{2}$QuantWare, Molengraaffsingel 8, 2629 JD Delft, The Netherlands}
\affiliation{$^{3}$Department of Physics, ETH Zurich, 8093 Zurich, Switzerland}
\affiliation{$^{4}$Quantum Center, ETH Zurich, 8093 Zurich, Switzerland}
\affiliation{$^{6}$Quantum and Computer Engineering Department, Delft University of Technology, 2628 CD Delft, The Netherlands}
\affiliation{$^{7}$QuTech, Delft University of Technology,  2628 CJ Delft, The Netherlands}

\date{\today}

\begin{abstract}
Modular integration has emerged as the main pathway for scaling superconducting quantum processing units (QPUs) beyond the constraints of fabrication yield and physical footprint. Currently, two primary strategies lead this effort. Mirroring the "Scaling Up" and "Scaling Out" approaches in GPU architectures and AI infrastructures, these are: chiplet-based scaling, which preserves dense connectivity and high gate fidelity at the expense of engineering complexity, and distributed architectures, which decouple system scaling from monolithic QPU advancements at the expense of sparser connectivity and lower interconnect quality. To evaluate these approaches, we introduce a quantitative stress test measuring the execution cost of a dense workload of random logical entangling operations using a surface code scheme. Using a dedicated compiler, we compute the space-time cost as the number of network nodes increases, analysing this scaling behaviour across various surface code distances, Bell-state fidelities, and Bell-pair generation times. We find that distributed architectures incur an up to exponential space-time performance penalty compared to an effectively monolithic architecture across all simulations. Our results also show that as the network grows, this penalty manifests in two distinct scaling regimes: a noise-dominated regime constrained by insufficient Bell-state fidelity and generation rates, and a connectivity-dominated regime bottlenecked by lattice-surgery routing congestion.
\end{abstract}

\maketitle

\section{INTRODUCTION}
Quantum computing offers a theoretical advantage in solving complex problems that are intractable with classical computing resources \cite{lloyd1996universal, Shor_1997,farhi2001quantum,aspuru2005simulated,arute2019quantum, alexeev2021quantum,google_2025_nmr}. However, successful execution of quantum algorithms requires hundreds to thousands of faultless logical qubits and billions of high-fidelity gate operations \cite{Dalzell_2025, beverland2022assessingrequirementsscalepractical}. To date, no physical qubit platform is inherently capable of meeting these demands, as they are fundamentally constrained by decoherence, high susceptibility to noise, and operational errors. Consequently, all practical quantum computing setups must rely on quantum error correction (QEC) to achieve the necessary fidelities by encoding arrays of physical qubits into robust logical qubits, which are then actively protected against errors through repeated syndrome measurement cycles~\cite{gottesman1997_thesis,fowler2012surface}. This comes at a substantial cost: achieving application-level logical error rates with QEC demands orders of magnitude more physical qubits than logical qubits, with estimates for early fault-tolerant workloads involving hundreds of thousands to millions of physical qubits~\cite{google_rsa, Zhou_2025,webster2026pinnaclearchitecturereducingcost}.

Among the candidate platforms striving to meet the demanding requirements of QEC, superconducting qubits have emerged as a leading modality, offering distinct architectural advantages such as high gate speeds and compatibility with established microelectronic fabrication methods~\cite{fang2026bridgingsuperconductingneutralatomplatforms,kjaergaard2020superconducting, jiang2025advancements}. Building on these foundations, superconducting platforms have demonstrated significant improvements in coherence time, gate fidelity, and readout fidelity in recent years~\cite{tuokkola2025methods, google2025quantum, gao2025establishing, abdurakhimov2024technologyperformancebenchmarksiqms}. However, in spite of these advancements, present-day superconducting quantum processing units (QPUs) have stalled in the hundred-qubit regime~\cite{google2025quantum,kim2023evidence,gao2025establishing,abdurakhimov2024technologyperformancebenchmarksiqms}.
The primary bottleneck restricting the scaling of monolithic superconducting QPUs is the rapid decrease in fabrication yield as the chip footprint expands, as well as the large physical footprint of the required input and output (I/O) lines. To overcome these limitations, two distinct modular scaling strategies have emerged. These strategies parallel the "scale-up" and "scale-out" paradigms established in classical GPU and AI computing infrastructures \cite{barroso2013datacenter, hetherington2015memcachedgpu, zheng2022alpa}.

The first strategy (scaling up) circumvents the yield problem by connecting distinct, pre-characterised chiplets via short-range, coherent links to form an \textit{effectively monolithic} QPU~\cite{qw_whitepaper, Rahamim_2017, paradkar2025superconducting, yu2023through,mayer20253d} (as presented in Fig.~\ref{fig:dist_vs_mono}a). By using 3D integration techniques, such as vertically integrating chiplets with a carrier chip or interposer using through-silicon vias (TSVs) and flip-chip bonding, vertical routing furthermore solves traditional I/O bottlenecks by alleviating the need for lateral routing. This also allows the system to maintain a continuous 2D lattice with nearest-neighbour coupling. 

The second strategy (scaling out) instead connects distinct QPUs into a distributed network. In this architecture, modules are linked via coherent microwave~\cite{ibm_long_range, Magnard2020} or optical~\cite{van2025optical} interconnects, either within the same cryostat or across multiple cryostats (illustrated in Fig.~\ref{fig:dist_vs_mono}c). These networks rely on inter-module Bell pairs to transfer quantum states and mediate nonlocal operations via teleportation-based primitives~\cite{caleffi2024distributed}. This distributed approach has gained significant traction and now features prominently in the roadmaps of various quantum computing companies~\cite{IBMQuantumRoadmap2026, IQMRoadmap2026, RigettiAFRL2024}.

However, both modular approaches face severe and distinct engineering challenges. For the effectively monolithic approach, the chip-to-chip interconnects must be nearly indistinguishable from on-chip elements. Early work shows entanglement rates and fidelities approaching strictly on-chip operations~\cite{norris2025_chiplet, gold2021_chiplet}, but these rates and fidelities need to be maintained at scale. In addition, the system must manage a rapidly growing overall I/O density, signal fanout, and increasing thermal loads. Scaling this dense 3D architecture is highly dependent on ongoing developments in heat load reduction~\cite{krinner2019engineering, raicu2025cryogenic, van2025optical, monarkha2024equivalence} and advanced cryogenics~\cite{MaybellColdCloud2026, bluefors2025kide}.

The distributed approach, on the other hand, uses inter-node links, which have significantly worse performance compared to their intra-node counterparts. Because logical operations must cross seam interfaces, the links must provide Bell pairs with sufficiently high fidelity, generation rate, and multiplexing capacity to avoid becoming the dominant source of logical error and latency. Achieving this is non-trivial; while microwave-based links are relatively well-studied, they face severe physical challenges in scaling multiplexing capacity and maintaining low photon loss over macroscopic distances~\cite{Axline2018, Campagne2018, Kurpiers2018, Heya2025, Qiu2025}. In contrast, optical interconnects natively support higher multiplexing abilities and negligible transmission losses over metre scales~\cite{gianini_silicon_2026}, but integrating them with superconducting circuits requires microwave-to-optical transduction, a nascent technology currently dominated by probabilistic and noisy conversion processes~\cite{sekine_microwave--optical_2025, mueller_high-rate_2024}. In addition to the quality of the links, proposed distributed architectures are also limited by their sparse inter-node connectivity ~\cite{niu2023low,ang2024arquin,jacinto2026network, yoder2025tourgross}. This forces inter-node logical operations to compete on a limited amount of possible routing paths, such that routing overhead and latency-induced idling errors can quickly dominate the algorithm's execution runtime~\cite{croot_enabling_2025,almanakly2025deterministic,marqversen2025fault}.

However, rather than viewing these strategies as strictly competing, realising million-qubit systems might require a complementary approach: making effectively monolithic nodes as large as possible and networking these nodes together into a large modular system. Previous work on such hybrid architectures has analysed fault-tolerant inter-node links, distributed lattice-surgery protocols, and network requirements for modular quantum computation~\cite{ramette2024fault,jacinto2026network,marqversen2025fault,pattison2024fast}. These studies establish that noisy Bell pairs can be incorporated into logical operations, but do not determine the actual cost of executing a compiled workload on a modular layout. A realistic evaluation must therefore include the full execution cost after logical compilation, as well as realistic noise modelling, including Bell pair modelling. To this end, we introduce a compilation-based quantitative stress test that maps Bell-pair specifications, QPU node sizes, node counts, code distance and physical noise to the resulting space-time execution overhead. This allows us to compare distributed integration to effectively monolithic scaling accurately.

\section{Modular fault-tolerant architectures and physical interconnect models}\label{sec:previous_work}

In this section, we establish the framework for comparing effectively monolithic and distributed quantum architectures. We first outline the fault-tolerant logical operations required in a modular surface-code layout. We then define the physical interconnect models needed to execute distributed lattice surgery, establishing the parameter space of Bell-pair generation rates and fidelities that govern network performance.

\subsection{Fault-tolerant logical gates in modular surface-code architectures}
Fault-tolerant quantum computation encodes quantum information in logical qubits to prevent the uncontrolled spread of physical errors. Throughout this paper, we use the surface code to define these logical qubits, which is a prominent scheme for superconducting architectures due to its high error threshold and natural mapping to nearest-neighbour hardware connectivity \cite{kitaev2003fault, fowler2012surface}. Specifically, we employ a distance-$d$ rotated surface code, which encodes one logical qubit into $2d^{2}-1$ physical qubits and produces logical error rates that decrease exponentially in $d$ when operating below the error threshold \cite{bravyi1998quantum}.

\begin{figure*} \centering \includegraphics[width=\textwidth]{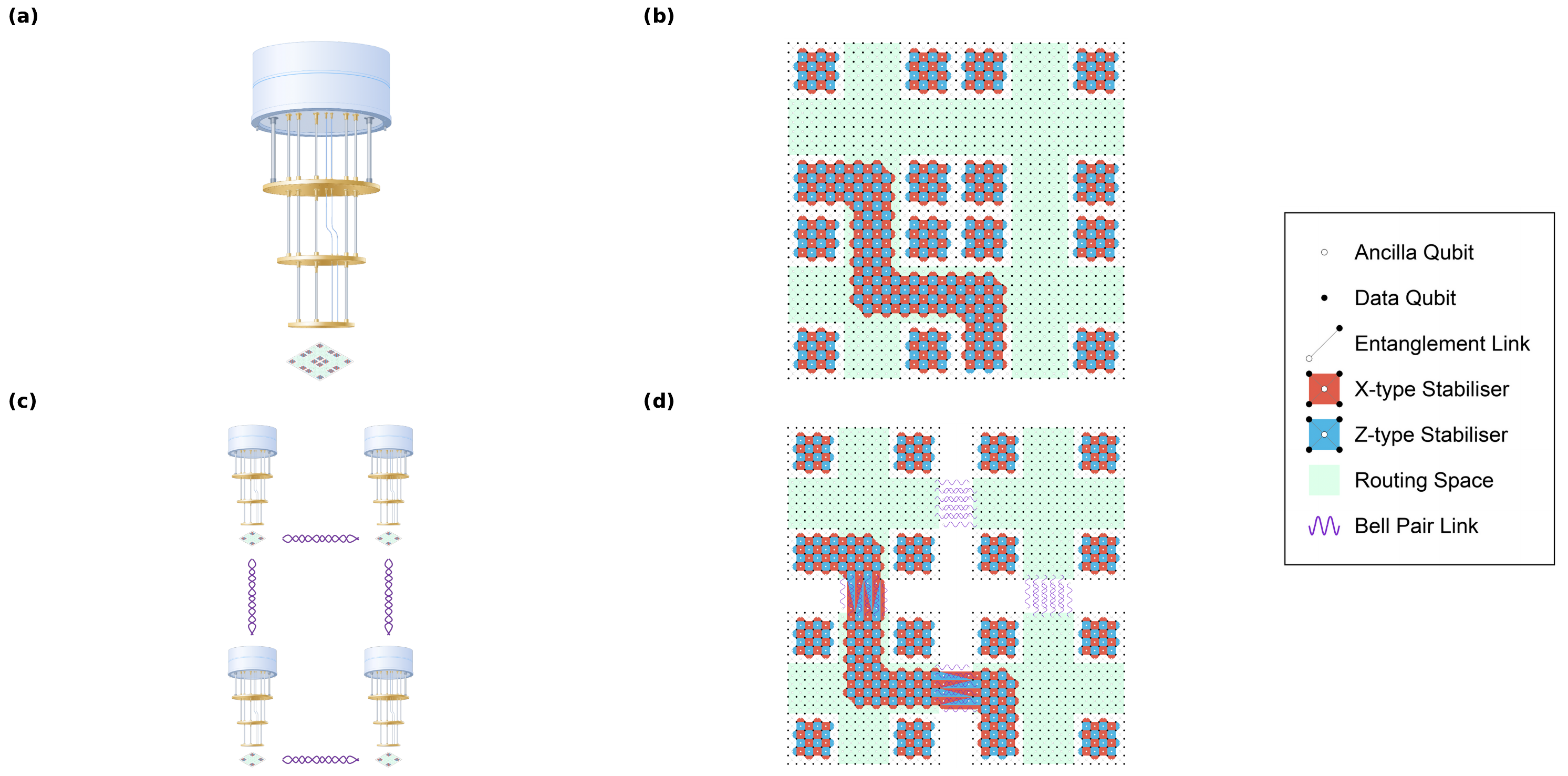} \caption{(a) In an effectively monolithic architecture, chiplets are connected to form a contiguous 2D layout. This design ensures the system operates as a single, unified QPU housed within a singular cryostat. This is analogous to "scaling up," achieved by integrating multiple chiplets to construct a single, high-capacity QPU within a single cryostat. (b) A joint parity measurement between two logical surface code patches across an effectively monolithic layout, using conventional lattice surgery. (c) In a distributed architecture, multiple smaller nodes (that may or may not be housed in separate cryostats) are connected using microwave or optical links, over which information is transferred. This is analogous to "scaling out," where distinct QPUs act as networked nodes, potentially distributed across multiple cryostats. (d) A joint parity measurement between two logical surface code patches on two separate nodes in a nearest-neighbour distributed connectivity layout, using seam-mediated lattice surgery. The seam-mediated stabiliser operations follow the procedure of Ref.~\cite{jacinto2026network}.}\label{fig:dist_vs_mono} \end{figure*}

In addition to information encoding, a universal set of fault-tolerant logical operations is needed to execute quantum algorithms on the logical qubits. Logical single-qubit Clifford operations using the surface code are commonly implemented through Pauli-frame tracking, and non-Clifford gates are typically realised by magic-state injection followed by distillation or cultivation \cite{10.1145/3061639.3062300, Bravyi_2005, lao2022magic, gidney2024magic}. However, multi-qubit Clifford gates, most notably the logical $\mathrm{CNOT}$, require fault-tolerant interactions between spatially separated logical qubit patches in the 2D lattice. In a monolithic layout, these operations are performed using lattice surgery \cite{lattice_surgery1,litinski2019game}, where code patches are temporarily merged and split by extending stabiliser measurements across a shared boundary for multiple rounds of syndrome extraction, as illustrated in Fig.~\ref{fig:dist_vs_mono}(b). These operations require dedicated ancillary patches between logical data patches known as routing regions (see Fig.~\ref{fig:dist_vs_mono}(b,d)), which consume additional space-time resources. Because the routing regions must remain available throughout the operation and cannot be shared in the same timestep, lattice surgery requires precise scheduling to avoid geometric conflicts. Such routing requirements increase circuit depth and prolong idle periods for other logical qubits, thereby increasing the probability of decoherence noise \cite{herr2017optimization,litinski2019game,hamada2024efficient}.

On a (effectively) monolithic chip, the continuous 2D layout provides ancillary routing regions that can bridge neighbouring code patches. However, in distributed architectures, this continuity is interrupted because nodes are connected via seam interfaces \cite{niu2023low,ang2024arquin}, forcing non-local logical interactions to funnel through a limited number of pathways. 
This exacerbates routing congestion and introduces additional idle time for the creation of entanglement between qubits. Moreover, operations across seams are typically slower and noisier than bulk stabiliser measurements, which affects both the logical operation time and decoding accuracy.
Several methods exist for implementing such inter-node logical gates, including teleported transversal gates, distilled logical links, and distributed lattice surgery \cite{marqversen2025fault,pattison2024fast}. 
In what follows, we focus on the latter, as this method naturally extends the monolithic lattice-surgery framework and avoids the spatial overhead of distillation protocols. 

In distributed lattice surgery, the continuous multi-qubit parity measurements of the monolithic approach are replaced by entanglement-mediated checks. The protocol consumes inter-node Bell pairs to effectively teleport the required stabiliser information across the network seam, as depicted in Fig.~\ref{fig:dist_vs_mono}d. This decomposes the non-local operation into local measurements at each node \cite{ramette2024fault,jacinto2026network}. For a boundary of length $d$, merging and splitting the patches requires generating and consuming $2d$ Bell pairs during every QEC cycle \cite{jacinto2026network, shalby2025optimized}.

\subsection{Physical constraints and future projections for superconducting Bell links}

The resulting logical performance is jointly constrained by the decoder's capabilities and the seam's hardware specifications, because these $2d$ inter-node Bell pairs are used to mediate the effective seam-syndrome extraction circuit. The most relevant hardware capabilities are the Bell-state fidelity, the generation rate, and the multiplexing capacity. Imperfect entanglement injects errors directly into the seam stabilisers, degrading the logical error threshold. Simultaneously, if the network cannot generate pairs fast enough, the QEC cycles must stall to wait for entanglement, exposing the data qubits to additional decoherence. 

At the time of writing, the most mature mid- to long-range interconnect technology for superconducting qubits utilises itinerant microwave channels, which have demonstrated remote entanglement over distances ranging from centimetres \cite{Axline2018, Campagne2018, Kurpiers2018, Heya2025} to tens of metres \cite{Storz2023, Qiu2025}. Scaling these proof-of-principle demonstrations into fault-tolerant resources requires simultaneous advances across all three hardware axes. Current links based on 3D-waveguides between different cryogenic systems achieve deterministic Bell-state fidelities around 85$\%$ \cite{Storz2025,Kulikov2024}, primarily limited by photon loss in the quantum channel. Heralded schemes and purification protocols can avoid this loss, having already culminated in fidelities in the 92$\%$ to 94$\%$ range \cite{Burkhart2021,Teoh2025,Yan2022}. However, heralded schemes do so by trading speed for fidelity; currently, these schemes operate at 1 to 10 kHz, well below the ${\sim}1~\mathrm{MHz}$ cadence needed to prevent surface-code cycles from stalling. To avoid this, hardware-level photon loss must be reduced. Realistic innovations here project deterministic fidelities of up to 98$\%$ \cite{Makihara2025, Storz2023}. Supporting this outlook, a recent demonstration achieved a Bell-state fidelity of 94$\%$ using a low-loss, 64-metre microwave cable implemented in a single cryostat \cite{Qiu2025}. Without any heralding overhead, the Bell-pair generation rates can theoretically approach the inverse of the emission/absorption times and propagation delays, reaching limits around 250 ns \cite{Zhong2021, Kurpiers2018}. Furthermore, as each active seam requires $\mathcal{O}(d)$ Bell pairs per cycle, substantial multiplexing capacity is mandatory. While the development of parallelisation currently lags behind fidelity improvements, recent simulations of frequency-multiplexed protocols suggest that 60 parallel photon transmissions over a 5-metre waveguide is a realistic decade-scale target \cite{Penas2024}. To systematically evaluate the potential of distributed architectures to extend quantum scaling, we look beyond current experimental limitations and probe a forward-looking parameter space.

\section{Methods}\label{sec:methods}
To quantify the performance impact of seam-based logical operations, we develop a comprehensive noise-simulation and compilation framework that compares distributed architectures with an effectively monolithic baseline. The framework combines a novel logical compiler for topology-aware gate scheduling with a lattice-surgery-based CNOT noise model that captures errors arising from inter-node Bell links. Finally, we define a set of quantitative metrics to evaluate the comparative overhead of distributed execution.

\subsection{Logical compilation}

\label{sec:logical_compiler}
Explicit logical compilation is crucial to accurately capture the impact of the constrained connectivity inherent to seam-based distributed architectures. To that end, we introduce a logical compiler that translates a target quantum circuit into a physical-level execution schedule, strictly adhering to the spatial connectivity constraints of the underlying hardware topology. Our compilation pipeline operates in three sequential stages: qubit assignment, routing, and scheduling.

\begin{figure*} \centering \includegraphics[width=\textwidth]{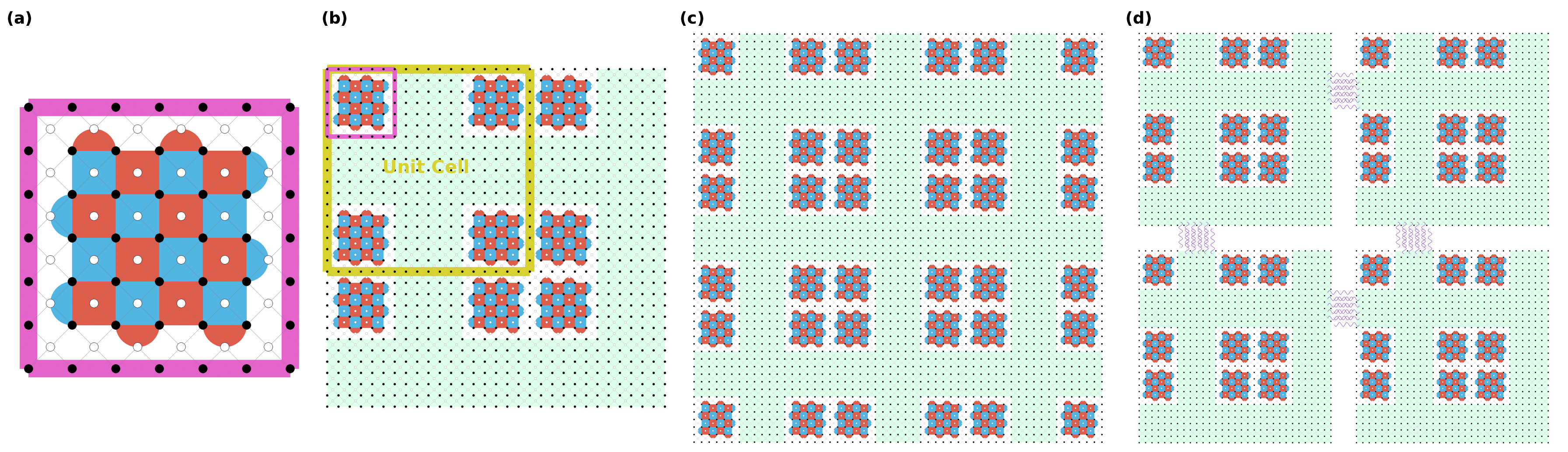} \caption{Tiling of our surface code unit cell to form logical architectures (a) A single distance $d$ logical qubit takes up a $(d+1)\times(d+1)$ lattice to allow for lattice surgery operations with directly neighbouring logical qubits. (b) Each unit cell contains nine such patches, out of which four are used as data qubits. This unit cell is then tiled to fill the lattice. We allow the unit cell to get cut off after the second column of the unit cell on the right of the lattice, or the second row at the bottom of the lattice.  In this way, routing space is reserved on the side to ensure the bottom-right data qubit is not cut off from lattice surgery operations. (c) An example of such tiling on an effectively monolithic chip. (d) An example of such tiling on a set of distributed nodes, where the nodes are connected via Bell pairs. All schematics include an outer qubit layer for visual clarity; these outer qubits are never used in lattice surgery operations and are therefore omitted from the physical qubit counts used in our calculations.
}\label{fig:unit_cell_tiling} \end{figure*}

In the qubit assignment stage, we allocate logical qubits to physical surface code patches within the available hardware lattice (Fig.~\ref{fig:unit_cell_tiling}a). Following the architecture of Chamberland and Campbell~\cite{chamberland2022universal}, we build this layout using a unit cell capable of hosting nine surface code patches of variable code distance $d$. Within this cell, four patches serve as logical data qubits at the corners, while the remaining five patches are dedicated to extending and merging code patches during lattice surgery (Fig.~\ref{fig:unit_cell_tiling}b). To optimise physical space at the edges of the lattice, unit cells may be truncated. However, we strictly enforce that truncation cannot leave any data patch isolated; every logical patch must retain access to both $X$-type and $Z$-type boundaries to ensure universal lattice-surgery operations. 

For effectively monolithic architectures, we tile these unit cells continuously across the entire lattice. Consequently, some routing boundaries may require inter-chiplet two-qubit gates to connect to the neighbouring routing space, as indicated in Fig.~\ref{fig:unit_cell_tiling}c. In the distributed setting, we assume identical, standalone nodes, meaning each node's lattice begins with a fresh unit cell boundary, as depicted in Fig.~\ref{fig:unit_cell_tiling}d.

During the routing stage, a routing algorithm maps each layer of the logical circuit under the geometric constraints of the topology to perform lattice surgery on the logical qubits. Given a layer of the circuit's (ideally parallelised) operations, the router identifies a conflict-free subset that can be executed without overlapping routing paths. Any remaining logical operations are deferred to subsequent circuit layers. Among feasible routing solutions, the compiler preferentially selects paths that minimise routing overhead and limit the number of additional circuit layers. In particular, unlike previous approaches that rely on evaluating fixed routing patterns~\cite{jacinto2026network}, our compiler explicitly determines dynamic routing paths for every logical operation across both architectures.

Once all logical operations are routed, the compiler enters the scheduling phase. Here, we evaluate each compiled circuit by determining the logical error rates for each operation. To quantify the cost of the entire architecture, we calculate an expected \emph{space-time} cost \cite{jacinto2026network}, $C$, defined as

\begin{equation}
C \;\equiv\; N_{\mathrm{phys}} \times T_{\mathrm{succ}},
\end{equation}

where $N_{\mathrm{phys}}$ denotes the total number of physical qubits required to instantiate the full fault-tolerant layout (including data patches, ancillary patches, routing space, and the Bell-pair seam qubits in the distributed configuration), and $T_{\mathrm{succ}}$ represents the expected time-to-first-success of the logical workload. The choice of $C$ stems from the fact that distributed architectures will inevitably consume a larger \textit{total} physical qubit count across all nodes to compensate for noisy seams and potentially more routing overhead.
We compute $T_{\mathrm{succ}}$ in three steps:
\begin{enumerate}
    \item Determine the wall-clock duration of a single full execution attempt from the compiled circuit as a function of the number of QEC rounds required for the chosen code distance and the additional decoder latency for the routed lattice-surgery operations. 
    \item Compute the \textit{success probability per attempt} by combining the success probabilities of all compiled logical operations. This accounts for lattice surgery errors, idling errors on non-participating patches, path-dependent additional decoding time, and waiting times during Bell-pair generation. 
    \item Convert this into an expected \textit{time-to-first-success} by dividing the single-attempt duration by the per-attempt success probability, which mathematically models the time required to achieve one successful run when repeating independent attempts.
\end{enumerate}
Finally, we evaluate $C$ across a sweep of code distances to identify the minimum-cost operating point for each architecture configuration.

\subsection{Numerical simulation framework}\label{sec:experimental_set_up}

To isolate the performance impact of inter-node connectivity, we use a random logical CNOT workload. This choice is intentional: random CNOT layers rapidly generate an effectively dense interaction graph, removing any locality or qubit placement advantages that would otherwise make the comparison strongly workload-dependent. Therefore, the workload serves as a connectivity stress test of the dominant architectural distinction between effectively monolithic and distributed layouts by demonstrating how efficiently entanglement operations can be scheduled or routed. In what follows, we present a high-level summary of the numerical framework; the complete theoretical model and derivation of the logical error rates for distributed CNOTs are detailed in App.~\ref{app:cnot}.

To physically model the execution of these logical CNOTs, we employ a standard lattice-surgery protocol. Specifically, we use an ancillary surface code patch initialised in the logical $\ket{+}$ state for each CNOT operation. We use lattice surgery to perform logical $ZZ$ and $XX$ measurements between the ancillary patch and the control or target patches, respectively. To execute CNOTs between logical qubits on separate nodes, the lattice-surgery boundary is extended across the inter-node interface via Bell pair seams, requiring $2d$ physical Bell pairs per traversed seam. For the $\text{M}_{ZZ}$ and $\text{M}_{XX}$ operations, we follow the theoretical model used by Jacinto \textit{et al.}~\cite{jacinto2026network} and treat both lattice surgery operations as rectangular memory patches over the number of stabiliser measurement steps required to perform them.

To quantify the logical error rates linked with these operations, we use a hardware-realistic noise model (described in App.~\ref{app:noise_model} and App.~\ref{app:noise_hardware}) to numerically estimate logical error rates for small surface-code memory patches, both with and without inter-node seams. Following the methodology presented in App.~\ref{app:fitting_model}, we fit an analytic surrogate model against the simulated data as a function of code distance and seam quality. We use this fit to extrapolate logical error rates for larger patches and long-range seam-mediated lattice-surgery operations beyond the capacity of Monte-Carlo simulations. 

In addition, the influence of idling is included directly in the numerical memory simulations, where we use the physical qubit idling error model from Ref.~\cite{marton2025optimal}. This is especially critical when the QEC cycle time and Bell-pair generation time do not match, as the resulting idling noise can quickly become a dominant error source. To calculate the workload's success probability, we compute the joint logical success probability of all participating CNOTs and idling logical patches, while tracking the shape and length of the path, which directly influence the logical error rates of lattice surgery. We further extend the model of Jacinto \textit{et al.}~\cite{jacinto2026network} by precisely tracking code distances and idling times throughout the circuit, and by adopting a more appropriate number of stabiliser cycles prior to syndrome decoding, allowing time-induced errors to be reliably neglected (App.~\ref{app:cnot}).

\subsection{Architectural constraints and workload}
To evaluate these architectures under realistic algorithmic and hardware constraints, we target the requirements of early fault-tolerant applications projected to operate in the MegaQuOp regime---requiring on the order of $10^6$ coherent logical operations~\cite{Preskill_2025}.

To establish our effective monolithic baseline, we sweep over admissible layouts and select the code distance \(d\) that minimises the expected qubit-time cost \(C\) for the target workload. This differs from choosing \(d\) from a standalone surface-code memory error rate alone. Logical memory experiments are widely used as a proxy for threshold behaviour and idling error suppression, but they do not capture the full error model of a compiled lattice-surgery workload ~\cite{google2025quantum,domokos2024characterization,litinski2019game}. Selecting \(d\) via the workload-level cost function $C$ also considers the combined trade-off between physical qubits, execution time, and total success probability across all \(10^6\) logical CNOT operations.

The admissible effectively monolithic layouts are discrete because the architecture is assembled from unit cells with dedicated routing space, as shown in Fig.~\ref{fig:unit_cell_tiling}. For each candidate code distance, we first determine how many distance-$d$ rotated surface code patches can fit within a single chiplet with a \(10^4\) physical qubit capacity, including the additional qubits needed between patches. Boundary routing regions that lie on the outer edge of the layout are not counted, since they are not needed for lattice-surgery access to another patch. Applying this construction to a \(10\times10\) chiplet grid identifies the minimum-cost baseline at \(d=27\), which fits exactly \(169\) logical qubits and requires \(656{,}100\) physical qubits. The corresponding one-shot success probability for the compiled workload is \(91.82\%\).

To ensure a fair and consistent comparison when evaluating the distributed architectures, we follow the steps below:
\begin{enumerate}
    \item \textbf{Fixed Workload:} We execute the exact same $10^6$ random logical CNOT operations on 169 logical qubits.
    \item \textbf{Node Capacity Limit:} No individual distributed node may exceed $N_\mathrm{phys, max} = 10^6$ physical qubits, reflecting the same underlying manufacturing constraints as the monolithic baseline.
    \item \textbf{Topology Sweep:} We distribute the 169 logical qubits across identical square nodes, yielding a feasible set of node counts $N_{\text{node}} \in \{3, 4, 7, 11, 19, 43, 169\}$ under the unit cell tiling constraints.
    \item \textbf{Code Distance Sweep:} Because distributed seams introduce additional noise, we allow the compiler to sweep the code distance $d \in [27,59]$ to find the new optimal balance between error suppression and hardware overhead for each node configuration.
\end{enumerate}

We note that at the logical execution level, we do not optimise the initial placement of logical qubits to minimise lattice surgery distance. Because we evaluate a random dense circuit workload, the qubit interaction graph rapidly mixes and, consequently, any local placement advantage becomes irrelevant after only a few layers of execution~\cite{keskin2025lattice}. Furthermore, at the network level, distributed nodes are arranged in a 2D nearest-neighbour grid. The feasible node counts ($N_{\text{node}}$) correspond directly to the discrete ways of tiling our 169 logical qubits across identical square nodes.\footnote{The configuration $N_{\text{node}}=169$ is an exception. Here, each node hosts a single logical data qubit. This node is instantiated with a minimal square logical layout comprising one data patch and three auxiliary logical patches to accommodate lattice surgery and Bell-pair interfaces. This plausibly represents the earliest practical stage of networked surface-code-based superconducting systems.} When $N_{\mathrm{node}}$ does not allow for a perfect square grid of modules, we use a partially filled square grid. Such incomplete grids feature fewer links and less path diversity, thereby increasing routing congestion and the serialisation of lattice-surgery operations. These limitations are less pronounced for larger $N_{\mathrm{node}}$, where increased path diversity mitigates routing congestion.

We investigate Bell-state fidelities ranging from 92$\%$ to 99$\%$, and generation times $t_\mathrm{Bell}$ from $250\,\mathrm{ns}$ to $2\,\mathrm{\mu s}$. We further assume that sufficient multiplexing capacity is available to simultaneously generate and use the \(2d\) Bell pairs required for each active seam connection per QEC cycle. By evaluating this parameter range, we aim to identify the threshold regimes where distributed networking becomes a viable strategy to complement and scale beyond the physical limits of individual monolithic QPU nodes.

\section{RESULTS}\label{sec:results}
In this section, we compare the compiled space-time cost of various distributed configurations against our effectively monolithic baseline. Specifically, we examine how interconnect performance, defined by Bell-pair generation time ($t_{\mathrm{Bell}}$) and fidelity ($F_{\mathrm{Bell}}$), dictates the execution cost. We map these operational boundaries under the aforementioned hardware setup and workload and determine the threshold specifications.

\subsection{The impact of interconnect performance on space-time cost}\label{sec:interconnect_performance}

\begin{figure*}
\centering
\centering
\begin{minipage}[tbh!]{\textwidth}
\includegraphics[width=\textwidth]{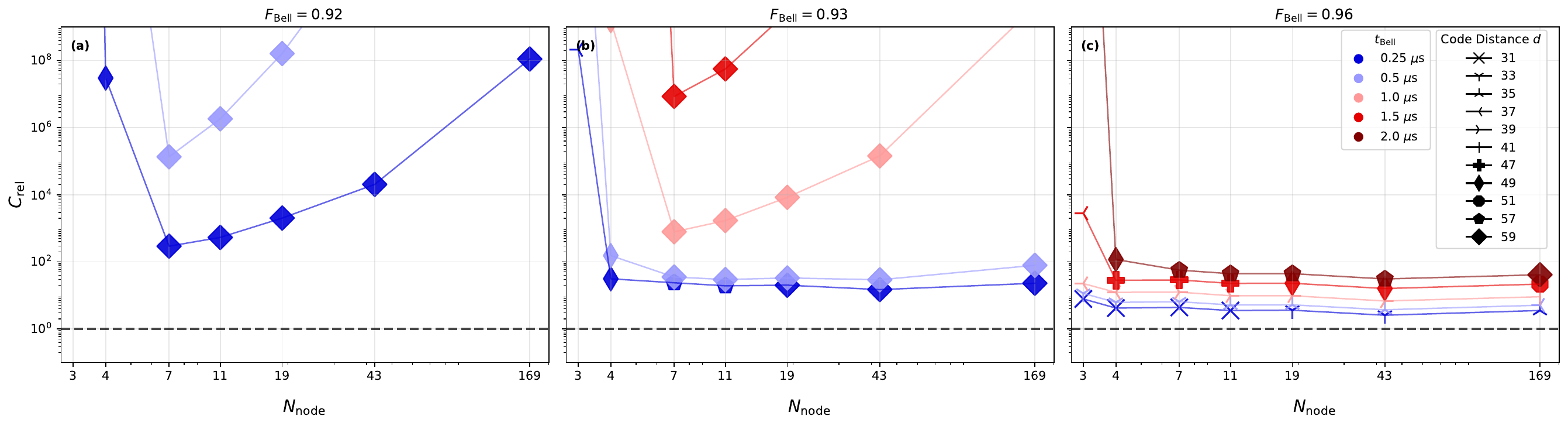}
\caption{\textbf{Dependence of cost on Bell-pair generation time $t_{\mathrm{Bell}}$.} Relative costs across the considered node counts for three fixed fidelities: (a) $92\%$, (b) $93\%$, and (c) $96\%$. The black dashed line indicates where the costs of computation for effectively monolithic and distributed architectures are equal.
Markers are coloured from blue to red to denote increasing $t_\mathrm{Bell}$.
Each marker has a different symbol depending on the code distance used.}
\label{fig:fixed_fidelity}
\end{minipage}
\end{figure*}

We evaluate the relative space-time cost ($C_{\mathrm{rel}} = C_{\mathrm{dist}}/C_{\mathrm{mono}}$) of the distributed architectures across a range of Bell-pair generation times and fixed fidelities, as shown in Fig.~\ref{fig:fixed_fidelity}. Across all evaluated regimes, the distributed approach carries a higher space-time cost than the effectively monolithic baseline ($C_{\mathrm{rel}} > 1$), showcasing an inherent overhead of network routing. However, the severity of this overhead and the optimal network size are heavily influenced by the physical performance of the interconnects.

In the low-fidelity regime ($F_\textrm{Bell} = 92\%$, Fig.~\ref{fig:fixed_fidelity}a), the distributed architecture is bottlenecked by seam-induced errors. The relative cost increases sharply, often superexponentially, as the network expands or as $t_\textrm{Bell}$ increases. At such low fidelities, the logical compiler is forced to select maximum code distances to compensate for the additional physical error channels introduced by the seams. Furthermore, expanding the network size in this regime is highly penalising: the requisite increase in inter-node seam traversals overwhelms any potential benefits from increased routing parallelism due to seam-induced errors.  

As fidelity improves to a transitional threshold ($F_\textrm{Bell} = 93\%$, Fig.~\ref{fig:fixed_fidelity}b), a clear bifurcation emerges based on the interconnect latency. For slower links ($t_\textrm{Bell} \ge 1\,\mu\mathrm{s}$), the interconnect latency approaches or exceeds the QEC cycle time. In these cases, idling errors accumulate on the data qubits while waiting for entanglement generation, compounding with the seam errors to drive exponential cost scaling. Conversely, for fast links ($t_\textrm{Bell} = 250\,\mathrm{ns}$), entanglement generation outpaces the QEC cycle time, mitigating idling errors. Here, the cost curves flatten as the number of nodes increases. This indicates a regime where the benefits of expanding the network, or specifically, resolving routing congestion and reducing the depth of the critical path, begin to effectively balance out the penalty of traversing the network seams.

For these lower Bell-state fidelities (e.g. \(95\%\) and below), the low node-count configurations (\(N=3,4\)) exhibit extremely high cost for all generation rates. This behaviour is not explained by routing congestion alone, but also by the node-size cap together with the discrete unit-cell tiling constraints described in Section~\ref{sec:methods}. Low-fidelity seams normally require larger surface-code distances \(d\) to suppress logical errors. However, at small $N_{node}$ values, 169 logical qubits must be packed into only a few modules, resulting in a higher logical qubit density per node. Because we have put strict bounds on the individual node capacities, each node cannot physically accommodate the required increase in \(d\). This forces the code distance to saturate prematurely at an insufficient size, hence resulting in a regime where the circuit both (i) stretches in time due to serialised routing and (ii) accumulates high per-cycle logical errors from repeated seam usage.

In the high-fidelity regime ($F_\textrm{Bell} = 96\%$, Fig.~\ref{fig:fixed_fidelity}c), seam fidelity is no longer the dominant bottleneck; rather, the execution cost is governed primarily by the Bell-pair generation time. In this regime, the overall space-time costs are significantly lower, and the curves remain relatively flat across different node counts. Furthermore, the combination of high fidelity and fast links allows the compiler to achieve the target logical error rate using smaller code distances (indicated by the marker shapes in Fig.~\ref{fig:fixed_fidelity}). As the node count increases and routing bottlenecks are alleviated, the compiler is able to drop the required code distance even further, as seen in the $t_\textrm{Bell} = 250\,\mathrm{ns}$ curve. 

\subsection{Distributed-to-monolithic relative cost across Bell-link specifications}\label{sec:heatmaps}

\begin{figure*}
\centering
\begin{minipage}[tbh!]{\textwidth}
\includegraphics[width=\textwidth]{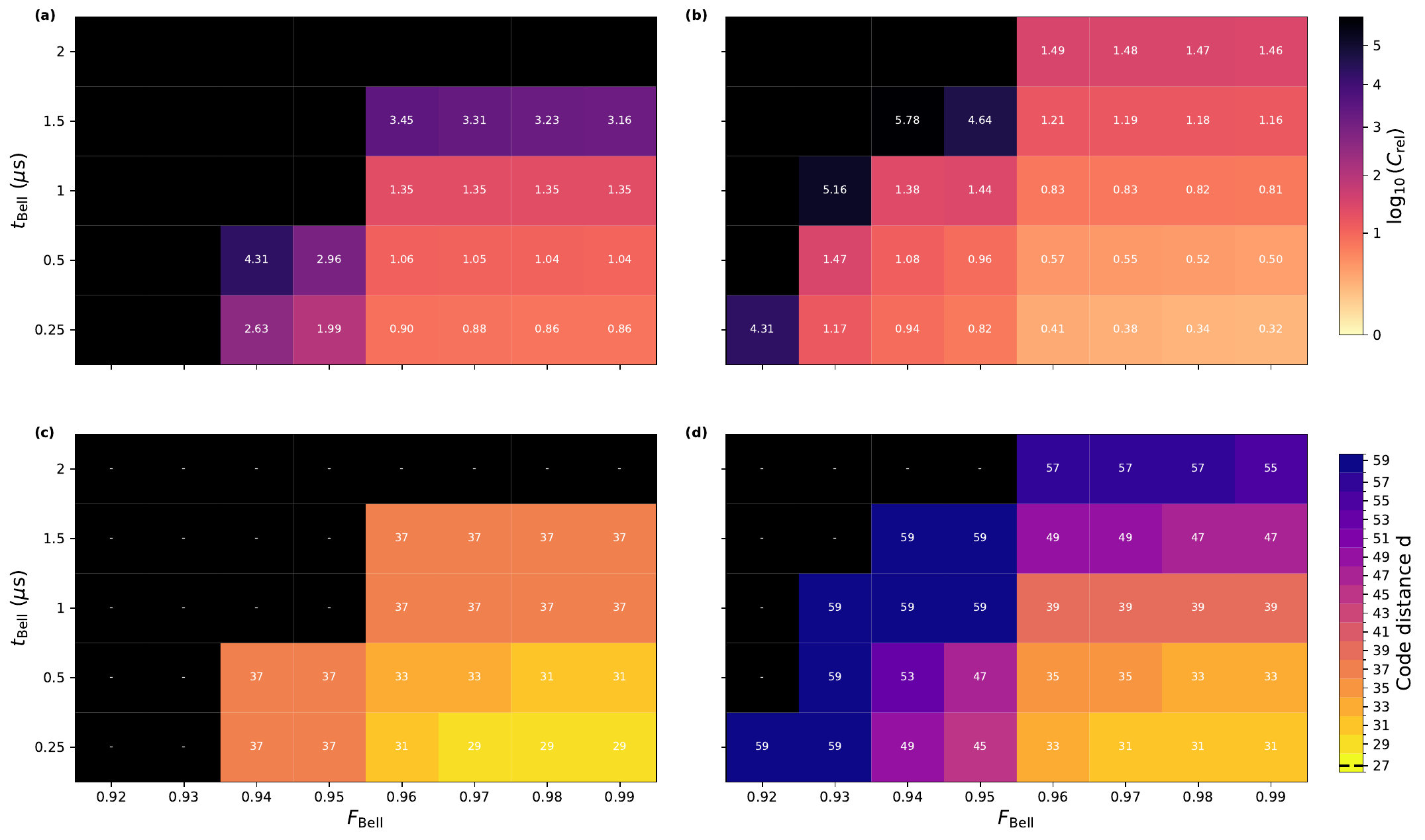}
\caption{\textbf{Cost and code distance scaling. Top panels (a, b):} Heatmap representation of the relative cost $C_\mathrm{rel}$ on a logarithmic scale as a function of Bell-pair fidelity and generation time $t_\mathrm{Bell}$ for (a) $N_\mathrm{node}=3$ and (b) $N_\mathrm{node}=43$. In black, we depict the regions where the relative cost is more than $10$ orders of magnitude higher than the baseline. \textbf{Bottom panels (c, d):} Optimal code distances utilised in panels (a) and (b), respectively. The fixed code distance required for the effectively monolithic baseline, $d=27$, is highlighted in the colourbar on the right.}
\label{fig:relcost_heatmaps}
\end{minipage}
\end{figure*}

To further illustrate the continuous interplay between entanglement rate and fidelity of Bell pairs, Fig.~\ref{fig:relcost_heatmaps} maps the relative space-time cost ($C_\mathrm{rel}$) across the complete 2D parameter space. We compare two distinct network configurations: a small network ($N_{\mathrm{node}}=3$, Fig.~\ref{fig:relcost_heatmaps}a), which represents a topologically constrained regime where routing constraints dominate, and a highly parallelised network ($N_{\mathrm{node}}=43$, Fig.~\ref{fig:relcost_heatmaps}b), which leverages large node counts to minimise routing congestion and lower the required code distance.

Both configurations exhibit a severe viability cliff: below specific rate and fidelity thresholds, the space-time cost of the distributed architecture becomes completely intractable (indicated by the dark regions), exceeding the effectively monolithic baseline by many orders of magnitude. Within the parameter space considered, the distributed architecture only approaches cost-parity with the monolithic baseline once Bell-pair fidelities reach approximately 96\%. At this threshold, the relative cost penalty narrows to roughly $0.5$ to $1.65$ orders of magnitude for the 3-node setup, and $0.3$ to $1.5$ orders of magnitude for the highly parallelised 43-node configuration. 

This sharp transition is physically corroborated by the optimal code distances required by the compiler, depicted in panels (c) and (d) of Fig.~\ref{fig:relcost_heatmaps}. In the long generation time, low-fidelity regimes, the compiler maximises the code distance ($d \to 59$) in a brute-force attempt to suppress seam-induced noise. Conversely, as fidelities cross the $96\%$ threshold and generation times drop, the requisite code distance rapidly relaxes toward the effectively monolithic baseline of $d=27$, visually mapping the exact parameter space where error correction regains its efficiency. We note that the $N_{\mathrm{node}}=43$ network requires a relatively higher $d$ to suppress the cumulative noise of its numerous seams, whereas $N_{\mathrm{node}}=3$ saturates at a lower code distance. In this small-size regime, performance is limited by routing congestion rather than noise, a structural bottleneck that brute-force increases in code distance cannot resolve.

These heatmaps confirm the findings of Sec.~\ref{sec:interconnect_performance}, showing a transition in the dominant performance bottleneck. Once the generation time $t_{\mathrm{Bell}}$ is equal to or less than the ${\sim}1\,\mu\mathrm{s}$ QEC cycle time, idling errors are effectively mitigated. Consequently, further improvements in interconnect speed yield diminishing returns. At this stage, the dominant lever for improving system performance is closing the fidelity gap between the distributed network seams and the local two-qubit operations.

Furthermore, we find that the networking penalty is not solely a function of Bell-link noise. To isolate the impact of connectivity constraints, we evaluated an ideal scenario in which Bell-pair specifications match those of local two-qubit gates ($30~\mathrm{ns}$ and $99.95\%$ fidelity), as detailed in App.~\ref{app:30ns}. Even in this connectivity-dominated regime, distributed architectures yield a relative cost penalty of $15$ to almost $400$ percent. This illustrates that even in this near-ideal scenario, the distributed architecture does not become fully equivalent to the effectively monolithic baseline for all node counts; a residual noise penalty always follows from the seam-dependent terms in the distributed lattice-surgery error model, as derived in App.~\ref{app:cnot}.

\section{DISCUSSION AND OUTLOOK}\label{sec:discussion_outlook}

\subsection{Discussion of results and implications}
Our results demonstrate that the execution cost comparison between effectively monolithic and distributed quantum processing units is defined by the physical properties of the quantum interconnects and the resulting topological constraints. Distributing a logical workload across multiple QPUs introduces an overhead characterised by three core effects: reduced routing freedom, lower fidelity operations during lattice surgery, and idling noise accumulated in every QEC cycle while waiting for Bell-pair generation. 

By evaluating a MegaQuOp-scale benchmark circuit with \(10^6\) random logical CNOT operations acting on 169 logical qubits, we establish physical bounds for this overhead. We observe that unless inter-node seams achieve a fidelity of $\geq 96\%$ and a Bell-pair generation rate of $\geq 1 \,\mathrm{MHz}$ (corresponding to $t_{\mathrm{Bell}} \le 1\,\mu\mathrm{s}$), the distributed architecture suffers an exponential increase in our space-time cost compared to the monolithic baseline. 

While the qualitative behaviour of these thresholds originates from the physical mechanics of the surface code, their exact numerical values are a product of our chosen modelling assumptions. Once entanglement generation significantly outpaces syndrome extraction, idling errors are effectively neutralised, yielding diminishing returns for further rate improvements. Likewise, under our assumed error model, the fidelity threshold represents the critical point at which seam-induced errors can be successfully suppressed without demanding overwhelmingly large code distances ($d \to 59$) that raise the hardware cost.

Our model assumes the availability of $2d$ parallel Bell pairs in every QEC cycle, consistent with the optimistic bandwidth assumptions made in distributed lattice-surgery resource estimates~\cite{jacinto2026network}. Constraining the parallel generation and consumption of Bell pairs per active seam to $m < 2d$ either constrains the admissible code distance effectively to \(d\le m/2\), or forces serialised seam operations over approximately \(\lceil 2d/m\rceil\) Bell-pair generation windows. The latter case increases the effective Bell-pair generation time, thereby increasing idling errors and shifting the viability thresholds reported here toward higher fidelities or faster links.

These constraints indicate an inherent threshold for modularity in near-term architectures that are based on the surface code. Because distributed systems require larger code distances to overcome seam-induced errors, nodes with only hundreds of physical qubits cannot offer sufficient real estate to host multiple large-$d$ logical qubits alongside the necessary routing space for lattice surgery. This suggests a minimum QPU size requirement: individual modules must scale significantly beyond current sizes before the increased qubit count of distributed architectures can outweigh the noise penalties of the seams.

\subsection{Outlook and future research}
The hardware specifications identified by our simulations present a significant engineering challenge for present-day superconducting interconnects, since Bell-pair generation has not been shown to simultaneously attain the rate and fidelity thresholds discovered in this work. In contrast, state-of-the-art multi-chiplet systems have already achieved entanglement rates and fidelities that nearly match strictly on-chip operations~\cite{norris2025_chiplet, gold2021_chiplet}. If the high connection quality of multi-chiplet systems can be retained in large-scale processors, effectively monolithic architectures appear to have a distinct performance advantage over networked architectures.

Several avenues exist to push the underlying hardware and architectural performance of interconnects beyond present limits. Within the microwave domain, promising development paths include implementing active reset mechanisms, mitigating channel losses, and increasing multiplexing, all of which can elevate Bell-pair generation fidelity without sacrificing generation rates. Should scaling these microwave capabilities prove too demanding, optical interconnects offer a natural alternative. However, while optical integration is rapidly advancing for modalities like colour centers~\cite{knaut_entanglement_2024, stolk_metropolitan-scale_nodate} and trapped ions~\cite{saha_high-fidelity_2025,liu2026long}, efficient microwave-to-optical transduction for superconducting circuits remains critically underexplored ~\cite{sekine_microwave--optical_2025}.

At the architectural level, advanced protocols such as encoding and distilling physical Bell pairs into logical Bell pairs \cite{pattison2024fast,marqversen2025fault} could relax physical rate requirements at the cost of additional ancillary space, potentially shifting the viability trade-off once practical implementations mature. Furthermore, the transition to quantum low-density parity check (qLDPC) codes offers a powerful path to reducing the massive physical-to-logical qubit overhead \cite{yoder2025tourgross, webster2026pinnaclearchitecturereducingcost, cross2025improvedqldpcsurgerylogical}. qLDPC codes may therefore ease the packing of higher-distance logical qubits into smaller physical nodes, although inter-module operations will still require ancillary routing structures and Bell-pair-mediated seams \cite{jacinto2026network}. Consequently, higher encoding rates benefit both monolithic and distributed architectures equally, without circumventing the fundamental latency and noise constraints of the network bottleneck. 

In turn, algorithm-specific structure allows for heuristic logical qubit assignment, where frequently interacting qubits are clustered within the same node to minimise seam usage. Because quantum circuit mapping is a computationally hard problem~\cite{steinberg2024lightcone, SpinQ,ArtA,beSnake}, heuristic outcomes depend heavily on the interplay between the circuit’s interaction graph \cite{medina_qmi_paper}, the device coupling constraints \cite{SpinQ, ArtA}, and the compiler optimisation objectives \cite{beSnake}. Future performance gains may be possible through locality-aware placement and increased scheduling flexibility, such as entanglement swapping to connect distant logical qubits across intermediate nodes~\cite{main2025distributed}. Our results establish a fundamental baseline using a fixed heuristic compiler, providing a necessary benchmark for future work investigating how advanced, workload-aware co-design can mitigate the networking penalties of distributed quantum architectures.

Finally, the impact of logical compilation and algorithm structure warrants further investigation. Our stress test used a random logical CNOT workload to provide a general benchmark of the architectural routing ability. In practice, algorithm structures vary widely. While some quantum algorithms feature highly localised interactions that minimise the need for inter-node communication, others—such as space-efficient implementations of RSA-2048 decryption \cite{google_rsa}—exhibit complex global dependencies and require gate densities far exceeding our benchmark. 

\section{Conclusion}\label{sec:conclusion}

In this work, we compared effectively monolithic and distributed superconducting quantum computing architectures by devising a compilation-based stress test to isolate their fundamental differences. To that end, we used a fixed workload of lattice-surgery-based logical operations and a dedicated compiler to accurately capture the cost increase imposed by modularity. Our framework improves upon previous models by incorporating realistic physical error channels and strict topological routing constraints directly into the execution evaluations.

Our results demonstrate a systematic cost increase across all distributed architectures compared to the effectively monolithic baseline. We isolated three dominant mechanisms driving this increase, all inherent to repeated seam-mediated lattice surgery: Bell-pair fidelity, Bell-pair generation rate, and sparser inter-node connectivity. Network seams introduce error channels and force operation serialisation, which reduces the per-attempt success probability and can exponentially scale execution costs. Furthermore, because noisy seams demand higher code distances, individual nodes must be scaled to a minimum physical qubit capacity to accommodate the requisite logical qubits and routing space. Consequently, for a large fraction of the explored parameter space, adding more nodes to the network progressively degrades performance.

Within the evaluated framework, we have identified a threshold at which Bell-pair specifications become sufficient to suppress the errors induced by inter-node lattice surgery: a Bell-pair fidelity of $\geq 96\%$ and a generation rate of $\geq 1\,\mathrm{MHz}$. This threshold separates two distinct architectural scaling regimes:
\begin{itemize}
    \item \textbf{A failure-dominated regime:} Where seam noise and/or latency-induced idling errors exponentially drive up the computational cost.
    \item \textbf{A connectivity-dominated regime:} Where link quality is no longer the primary bottleneck, and routing congestion at the interconnects mainly dictates the system overhead.
\end{itemize}
This dichotomy suggests that as interconnect hardware matures, the primary bottleneck in distributed architectures will shift from a physical, noise-dominated regime to a topological challenge of connectivity and scheduling.

The performance thresholds required for distributed architectures to match an effectively monolithic baseline significantly exceed current technological capabilities. Therefore, our findings indicate that efforts to scale superconducting processors for classically intractable problems should prioritise the "scale-up" paradigm: maximising the physical qubit count within a highly integrated module. Once single-module limits are reached, further scaling will necessitate a transition to a "scale-out" approach, a strategy that becomes viable only alongside substantial improvements in inter-node seam quality.

\section{Acknowledgments}
We acknowledge Cesar Hernando de la Fuente, Ioana-Lisandra Draganescu, Edgar Roussel, and Qu Tianchen for their contributions to the development and implementation of the fitting procedure used in this work. We thank Joseph Gabriel Richardson for providing input on optical links as interconnect channels. We also thank Laura Sandoiu for the design of Fig.~\ref{fig:dist_vs_mono}.

S.S. was supported by the Swiss National Science Foundation and by the QuantERA II Program that has received funding from the European Union’s Horizon 2020 research and innovation program under grant agreement no 101017733.

\let\selectlanguage\relax
\bibliography{bibliography}

\appendix

\clearpage

\section{Noise model}\label{app:noise_model}

In this section, we present our physical noise model used to perform numerical simulations of surface code memory experiments. 
We use the following models for depolarising noise on one and two qubits, respectively:
\begin{equation}
\begin{split}
\mathcal{N}_\text{dep1}(\rho,p_1) &= (1-p_1)\rho + \frac{p_1}{3}\sum_{P\in\{X,Y,Z\}}P\rho P^\dagger,\\
\mathcal{N}_\text{dep2}(\rho,p_2) &= (1-p_2)\rho + \frac{p_2}{15}\sum_{(P_1,P_2) \in \mathcal{P}^2} (P_1\otimes P_2) \rho (P_1\otimes P_2)^\dagger.
\end{split}
\end{equation}
with $\mathcal{P}^2 = \{\mathbb{I},X,Y,Z\}^2 \symbol{92} (\mathbb{I},\mathbb{I})$. 
These channels are used to model noise on one-qubit respectively two-qubit gates. 
Additionally, we use a crosstalk model~\cite{zhou2025surface} for two-qubit operations to emulate crosstalk associated with two-qubit gates on superconducting hardware:
\begin{equation}
\mathcal{N}_\text{ct}(\rho,p_\text{ct}) = (1-p_\text{ct})\rho + p_\text{ct} (Z\otimes Z) \rho (Z\otimes Z).
\end{equation}
With $\mathcal{N}_\text{gate}(\rho)$ the action of the noiseless gate on a state $\rho$, we apply these models as $\mathcal{N}_\text{dep1}(\mathcal{N}_\text{gate}(\rho),p_1)$ for one-qubit gates and $\mathcal{N}_\text{dep2}(\mathcal{N}_\text{ct}(\mathcal{N}_\text{gate}(\rho),p_\text{ct}),p_2)$ for two-qubit gates. We also take into account the duration of one-qubit and two-qubit gates with the parameters $t_1$ and $t_2$, respectively.

For qubit initialisation and qubit measurement errors, we make use of bit-flip errors (for initialisation and measurement in the $Z$-basis) and phase-flip errors (for the $X$-basis):
\begin{equation}
\begin{split}
\mathcal{N}_\text{bf}(\rho,p_\text{bf}) &= (1-p_\text{bf})\rho + p_\text{bf} X \rho X, \\
\mathcal{N}_\text{pf}(\rho,p_\text{pf}) &= (1-p_\text{pf})\rho + p_\text{pf} Z \rho Z.
\end{split}
\end{equation}
Specifically, we refer to measurement errors with $\mathcal{N}_\text{meas}(\rho,p_\text{meas})\equiv\mathcal{N}_\text{bf,pf}(\rho,p_\text{meas})$, and to initialition errors with $\mathcal{N}_\text{init}(\rho,p_\text{init})\equiv\mathcal{N}_\text{bf,pf}(\rho,p_\text{init})$. For both operations, we assign a read-out duration time $t_\text{ro}$. 

We model the decoherence of idling qubits over time using the idling model from Martón and Asbóth~\cite{marton2025optimal}:
\begin{equation}
\label{eq:idling_model_marton}
\begin{split}
\mathcal{N}_\text{idle}(\rho,T_1,T_2,t) &= p_0(T_1,T_2,t) \rho + p_x(T_1,t) X \rho X \\
& \quad + p_y(T_1,t) Y \rho Y + p_z(T_1,T_2,t) Z \rho Z, \\
p_x(T_1,t) &= p_y(T_1,t) = \frac{1}{4}(1-\text{e}^{-t/T_1}), \\
p_z(T_1,T_2,t) &= \frac{1}{2} (1-\text{e}^{-t/T_2})-\frac{1}{4} (1-\text{e}^{-t/T_1}), \\
p_0(T_1,T_2,t) &= 1 - p_x(T_1,t) - p_y(T_1,t) - p_z(T_1,T_2,t).
\end{split}
\end{equation}

Lastly, we model Bell pairs between nodes as noisy versions of $\ket{\Phi^+} = (\ket{00}+\ket{11})/\sqrt{2}$~\cite{jacinto2026network}:
\begin{equation}
\begin{split}
\rho_\text{Bell} &= \mathcal{N}_\text{dep2}\big(\ket{\Phi^+} \bra{\Phi^+},p_\text{Bell} \big) \\
&= \Big(1-\frac{3p_\text{Bell}}{4}\Big) \ket{\Phi^+}\bra{\Phi^+} \\
& \quad + \frac{4p_\text{Bell}}{15} \big(\ket{\Phi^-}\bra{\Phi^-} + \ket{\Psi^+}\bra{\Psi^+} + \ket{\Psi^-}\bra{\Psi^-}\big).
\end{split}
\end{equation}
We assign a time $t_\text{Bell}$ to the duration of generating a Bell pair.

\section{Hardware parameters}\label{app:noise_hardware}
An overview of the values used for the parameters in these error models can be found in Table~\ref{tab:noise_error_parameter_values}. 
Apart from the ``regular'' two-qubit gate between two qubits on the same node/chiplet (with error probability $p_2=0.0005$), our analysis technically also includes a second type of two-qubit gate between qubits that are part of different chiplets in the inter-chiplet architecture. 
In the surface code memory channel, these gates are only active in a one-dimensional sublattice on the ``seam'' between two chiplets. 
Therefore, in this scenario, the intra-chiplet ``bulk'' gates contribute to most of the error budget. 
It has been shown that the effect of a one-dimensional seam can be neglected in a situation where $p_\text{bulk}/p_\text{bulk}^*\gg p_\text{seam}/p_\text{seam}^*$, with $p_\text{bulk}$ and $p_\text{seam}$ the error probabilities of operations in the bulk and seam, respectively, and $p_\text{bulk}^*$ and $p_\text{seam}^*$ there respective threshold values, shown to be in the range of $p_\text{bulk}^*\sim 1\%$ and $p_\text{seam}^*\sim 10\%$~\cite{ramette2024fault}.

As initial work shows entanglement rates and fidelities of inter-chiplet gates approaching strictly on-chip operations~\cite{norris2025_chiplet, gold2021_chiplet}, we assume seam errors are equivalent to the bulk error probability of $p_\text{bulk}=p_2=0.0005$. In contrast, the seam error rates $p_\text{Bell}$ used for distributed architectures cannot be ignored using the same reasoning, as they are more than an order magnitude worse than $p_2$.

The values in Table~\ref{tab:noise_error_parameter_values} largely follow the forward-looking target parameters listed in Ref. \cite{mohseni2025buildquantumsupercomputerscaling}, with a few exceptions. $T_1$ and $T_2$ are both taken more optimistically, which benefits distributed architectures because of the chosen noise model described in Eq. \ref{eq:idling_model_marton}. No distinction between inter- and intrachiplet two-qubit gates has been made based on recent advancements in inter-chiplet gates \cite{norris2025_chiplet, gold2021_chiplet}. 
The varied Bell-state transfer parameters $p_\text{Bell}$ and $t_\text{Bell}$ are also chosen based on forward-looking parameters on the decade scale, as described in Sec.~\ref{sec:previous_work}.

\begin{table}[h]
\centering
\begin{tabular}{|c|c|}
\hline
Parameter & Value \\ \hline
$p_1$ & 0.0002 \\ \hline
$p_2$ & 0.0005 \\ \hline
$p_\text{ct}$ & $0.0006^{1/3}$ \\ \hline
$t_1$ & $20$ ns \\ \hline
$t_2$ & $30$ ns \\ \hline
$p_\text{meas}$ & 0.003 \\ \hline
$p_\text{init}$ & 0.0025 \\ \hline 
$t_\text{ro}$ & $200$ ns \\ \hline
$T_1$ & $300$ $\mu$s \\ \hline
$T_2$ & $250$ $\mu$s \\ \hline
$p_\text{Bell}$ & variable \\ \hline
$t_\text{Bell}$ & variable \\ \hline
\end{tabular}
\caption{Values used for noise error parameters.}
\label{tab:noise_error_parameter_values}
\end{table}

\section{Fitting method}\label{app:fitting_model}
In this section, we discuss the fitting procedure for three numerical simulations of surface code memories. We simulate a memory channel of the effectively monolithic square surface code for both logical $X$ and logical $Z$ errors, giving rise to fitting parameters $\alpha_x$ and $p_{\text{th},x}$, respectively, $\alpha_z$ and $p_{\text{th},z}$.
In addition, we simulate a memory channel of a square surface code patch with a Bell pair seam in the logical $X$ direction for logical $X$ errors, giving rise to fitting parameters $\alpha_{1,2,3,\text{c}}$, $p^*$, and $p_\text{Bell}^*$. 
We follow the same fitting procedure as Ref.~\cite{jacinto2026network}, using the physical noise model introduced in Appendix~\ref{app:noise_model}. 
We use Stim~\cite{gidney2021stim} to simulate the operations of the surface code channels, adopt PyMatching~\cite{higgott2022pymatching} as the error sequence decoder, and employ Lmfit~\cite{newville2016lmfit} to fit the numerical results with the analytical models introduced below and obtain statistical errors for the values of the fitting parameters. 

For the effectively monolithic simulation, we use the fitting model
\begin{equation}
\begin{split}
& f_{d,x}(p) = \alpha_x \bigg(\frac {p}{p_{\text{th},x}}\bigg)^\frac{d+1}{2}, \\
& f_{d,z}(p) = \alpha_z \bigg(\frac{p}{p_{\text{th},z}}\bigg)^\frac{d+1}{2},
\end{split}
\end{equation}
for logical $X$ and logical $Z$ errors, respectively. To determine the logical $X$ ($Z$) error probability, we follow these steps:
\begin{enumerate}
    \item Initialise the surface code patch in the logical $\ket{0}$ ($\ket{+}$) state
    \item Perform $k$ rounds of stabiliser measurements
    \item Decode the error syndrome with a minimum weight perfect matching decoder
    \item Correct the errors found
    \item Measure all qubits individually in the $Z$ ($X$) basis. 
\end{enumerate}
To reach the asymptotic per-round logical error regime, we use $k=d$ rounds of stabiliser measurements before decoding.
We collect $n_\text{MC}$ Monte Carlo samples, of which $n_{d,x}(p)$ and $n_{d,z}(p)$ contain a logical $X$ and logical $Z$ error, respectively. 
Depending on the parameter regime, we use up to $n_\text{MC}=10^9$ samples in order to reduce statistical uncertainty and improve the accuracy of the fitted logical-error model. 
In addition, we perform the fits separately over multiple $p_\text{Bell}$ intervals for each fixed $t_\text{Bell}$. 
This finer partitioning improves the accuracy of the fitted parameters across the full fidelity range. 
We obtain estimators for the logical error probabilities per stabiliser measurement round via $\hat{f}_{d,x}(p)=1-(1-n_{d,x}(p)/n_\text{MC})^{1/k}$ and $\hat{f}_{d,z}(p)=1-(1-n_{d,z}(p)/n_\text{MC})^{1/k}$.

For the distributed simulations, we use the following fitting model \cite{ramette2024fault}:
\begin{equation}
\begin{split}
& g_d(p, p_\text{Bell}) = \alpha_1 \bigg(\frac{p_\text{Bell}}{p_\text{Bell}^*}\bigg)^\frac{d+1}{2} + \alpha_2 \bigg(\frac{p}{p^*}\bigg)^\frac{d+1}{2} \\
& \quad + \alpha_3 \sum_{i=1}^d \bigg(\frac{p_\text{Bell}}{p_\text{Bell}^{**}}\bigg)^\frac{i}{2}\bigg(\frac{p}{p^*}\bigg)^\frac{d+1-i}{2},\\
&p_\text{Bell}^{**} = \frac{p_\text{Bell}^*}{1 + \frac{\alpha_\text{c}}{1-\sqrt{p/p^*}}}.
\end{split}
\label{eq:distributed_simulation_model}
\end{equation}
For the simulation fitted with Eq.~\eqref{eq:distributed_simulation_model}, we determine logical $X$ errors by initialising the patch in the logical $\ket{0}$ state and finally measuring all qubits individually in the $Z$ basis. 

In obtaining the fitted parameter values, we first vary the Bell-state fidelity with the parameter $p_\text{Bell}$ while keeping the other parameters in Table~\ref{tab:noise_error_parameter_values} fixed. 
Next to that, we vary noise parameters $p_1$, $p_2$, $p_\text{meas}$, and $p_\text{init}$ simultaneously with a common multiplier, while keeping the remaining parameters fixed. 
Therefore, when we mention a specific value for $p\equiv p_1$ in this paper, the noise parameters $p_2$, $p_\text{meas}$, and $p_\text{init}$ are rescaled as $p_2=0.005p/0.0002$, $p_\text{meas}=0.003p/0.0002$, and $p_\text{init}=0.0025p/0.0002$. 

To obtain a fitting model for fixed $p$ and $p_\text{Bell}$ (at a specific value for $t_\text{Bell}$), we individually vary both $p$ and $p_\text{Bell}$ for distances $d\in\{5,7,9,11,13\}$ in a small range of less than half an order of magnitude around the target value while keeping all other parameters fixed, and fit the theoretical model based on the probabilities of logical errors found for both variations. 
For example, to obtain fitting parameters for $p=1.5\cdot10^{-4}$ and $p_\text{Bell}=7.6\cdot10^{-2}$ with $t_\text{Bell}=250$ $n$s, we first vary $p$ for ten values in the range $p\in[1.4\cdot10^{-4}, 2.8\cdot10^{-4}]$ while keeping $p_\text{Bell}=7.6\cdot10^{-2}$ and $t_\text{Bell}=250$ $n$s fixed. 
Subsequently, we vary $p_\text{Bell}$ for ten values in the range $p_\text{Bell}\in[5\cdot10^{-2}, 8\cdot10^{-2}]$ while keeping $p=1.5\cdot10^{-4}$ and $t_\text{Bell}=250$ $n$s fixed. 
Finally, we use all the data obtained for the fit. In this specific example, this resulted in $\alpha_1=0.073(2)$, $\alpha_2=0.054(5)$, $\alpha_3=0.0100(7)$, $\alpha_\text{c}=0.291(1)$, $p^*=0.00120(3)$, and $p_\text{Bell}^*=0.236(3)$. 
The reported values have been rounded for presentation purposes. 
For low Bell-state fidelities and long Bell-generation times, the fits typically yield larger values of the reduced chi-square, $\chi_{\mathrm{red}}^2$; for example, in the case above $\chi_{\mathrm{red}}^2 \approx 87$.

To analyse the validity of our approach, we ran our fitting simulations using the same noise model and assumptions of Ref.~\cite{jacinto2026network}.
This resulted in $\alpha_1 = 0.096(8)$, $\alpha_2 = 0.049(4)$, $\alpha_3 = 0.054(2)$, $\alpha_\mathrm{c} = 0.19(3)$, $p^* = 0.00587(1)$, and $p^*_{\mathrm{Bell}} = 0.248(6)$, $p_{\text{th},xz} = 0.00740(3)$, and $\alpha_{xz}=0.0490(7)$.

\section{Theoretical CNOT model}\label{app:cnot}
\begin{figure}[] \centering \includegraphics[width=0.8\linewidth]{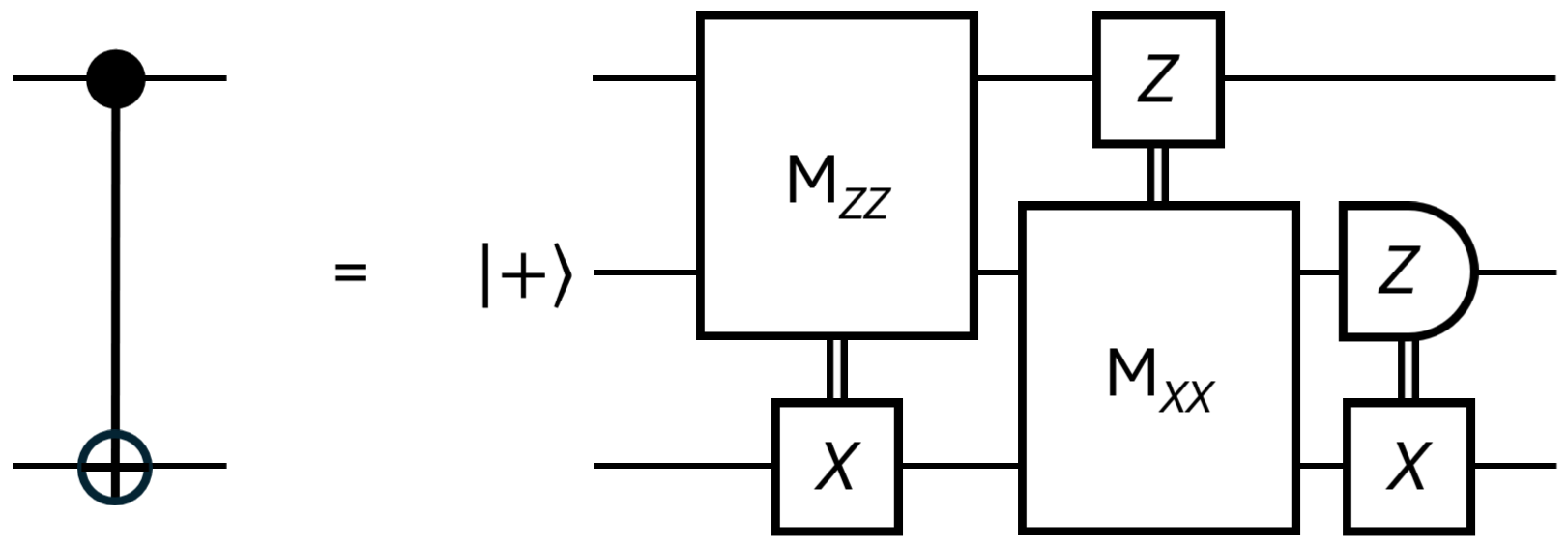} \caption{Circuit for a CNOT operation between logical surface code patches, with (planar) surface code patches of distance $d$ representing the qubits (horizontal lines in the diagram). The $\text{M}_{ZZ}$ and $\text{M}_{XX}$ operations indicate joint $ZZ$ and $XX$ measurements, respectively. Adapted from Ref. \cite{fowler2019lowoverheadquantumcomputation}.} \label{fig:cnot_figure} \end{figure}

In this section, we provide the theoretical derivation of error rates for logical CNOTs across a distributed architecture.
For each CNOT operation in our analysis, we use an ancillary surface code patch initialised in the logical $\ket{+}$ state with the circuit of Fig.~\ref {fig:cnot_figure}. 
In this configuration, the logical $ZZ$ ($XX$) measurement between the ancillary patch and the control (target) patch is performed using lattice surgery. For the $\text{M}_{ZZ}$ operation, we choose to use a surface code patch in the routing space, which is on the same node, directly next to the control patch of the logical CNOT. 
During the $\text{M}_{XX}$ operation, we then connect this ancillary patch in the vertical direction with the target patch of the CNOT. 
This target patch can be part of a different node than the control patch, in which case we use seams of $2d$ physical Bell pairs per seam to connect the patches.
This approach uses Bell pairs (instead of single ancillary qubits) to measure surface code stabilisers with data qubits on both sides of the seam. Specifically, we use the implementation introduced in Refs.~\cite{jacinto2026network} and \cite{shalby2025optimized} to realise these seam-based stabiliser measurements.

To model the $\text{M}_{ZZ}$ and $\text{M}_{XX}$ operations, we use the theoretical model derived and deployed in Jacinto \textit{et al.}~\cite{jacinto2026network}. 
This means both lattice surgery operations are modelled as rectangular memory patches over the number of stabiliser measurement steps required to perform these operations. 
For the $\text{M}_{ZZ}$ operation, we make use of the monolithic memory model of Ref.~\cite{jacinto2026network}, and approximate the total logical error probability $\mathbb{P}_{{ZZ},\text{rp}}$ by
\begin{equation}
\begin{split}
&\mathbb{P}_{{ZZ},\text{rp}} \approx P_{\text{L},1}^X + P_{\text{L},1}^Z,\\
&P_{\text{L},1}^X \equiv \alpha_x \bigg(\frac{p}{p_{\text{th},x}}\bigg)^{\frac{d_x+1}{2}}=\alpha_x \bigg(\frac{p}{p_{\text{th},x}}\bigg)^{\frac{2d+1}{2}},\\
&P_{\text{L},1}^Z \equiv \frac{d_x}{d_z}\alpha_z \bigg(\frac{p}{p_{\text{th},z}}\bigg)^{\frac{d_z+1}{2}}=2\alpha_z \bigg(\frac{p}{p_{\text{th},z}}\bigg)^{\frac{d+1}{2}}.
\end{split}
\label{eq:cnot_first_part_ZZ}
\end{equation}
In Eq.~\eqref{eq:cnot_first_part_ZZ}, $P_{\text{L},1}^{X,Z}$ describes the probability of a logical $X$ or $Z$ error, respectively. 
The parameter $p$ represents the value of the error probability on the physical level---e.g., the error probability of a two-qubit gate between two physical qubits. 
The parameters $\alpha_{x,z}$ and $p_{\text{th},x,z}$ are fitting parameters from numerically simulating a monolithic surface code memory patch with the specific physical error models describing the hardware---we discuss the noise model and fitting procedure in Appendix~\ref{app:noise_model} and Appendix~\ref{app:fitting_model}, respectively. 
The distances $d_{x,z}$ are defined as the minimum weights of the logical $X$ and $Z$ operators respectively. 
Lastly, following Ref.~\cite{jacinto2026network}, the factor $d_x/d_z$ is introduced in $P_{\text{L},1}^Z$ to approximate the effect of a higher logical error probability in the $Z$ direction as a result of extra vertical error strings that can occur along the prolonged horizontal dimension of the patch. 

For the $\text{M}_{XX}$ operation, we make use of the distributed memory model of Ref.~\cite{jacinto2026network}---or, the same monolithic model in case the control and target patch are part of the same node. 
Here, we approximate the total logical error probability $\mathbb{P}_{XX,\text{rp}}$ by
\begin{equation}
\begin{split}
&\mathbb{P}_{XX,\text{rp}}\approx P_{\text{L},2}^X+P_{\text{L},2}^Z,\\
&P_{\text{L},2}^X \equiv \begin{cases}
  P_{\text{L},2,\text{s}}^X & \text{if $n_\text{s}>0$,} \\
  P_{\text{L},2,\text{ns}}^X & \text{if $n_\text{s}=0$,}\end{cases}\\
&P_{\text{L},2,\text{s}}^X \begin{aligned}[t]
& \equiv n_\text{s} \alpha_1 \bigg(\frac{p_\text{Bell}}{p_\text{Bell}^*}\bigg)^\frac{d_x+1}{2} + \frac{d_z}{d_x} \alpha_2 \bigg(\frac{p}{p^*}\bigg)^\frac{d_x+1}{2} \\
& \qquad + n_\text{s} \alpha_3 \sum_{i=1}^{d_x} \bigg( \frac{p_\text{Bell}}{p_\text{Bell}^{**}} \bigg)^\frac{i}{2} \bigg(\frac{p}{p^*}\bigg)^\frac{d_x+1-i}{2} \\
& \leq n_\text{s}\alpha_1 \bigg(\frac{p_\text{Bell}}{p_\text{Bell}^*}\bigg)^\frac{d+1}{2} + n_\text{p} \alpha_2 \bigg(\frac{p}{p^*}\bigg)^\frac{d+1}{2} \\
& \qquad + n_\text{s} \alpha_3 \sum_{i=1}^d \bigg(\frac{p_\text{Bell}}{p_\text{Bell}^{**}}\bigg)^\frac{i}{2}\bigg(\frac{p}{p^*}\bigg)^\frac{d+1-i}{2},\\
&p_\text{Bell}^{**} \equiv \frac{p_\text{Bell}^*}{1 + \frac{\alpha_\text{c}}{1-\sqrt{p/p^*}}},\end{aligned}\\
&P_{\text{L},2,\text{ns}}^X \equiv \frac{d_z}{d_x} \alpha_x \bigg(\frac{p}{p_{\text{th},x}}\bigg)^\frac{d_x+1}{2} \leq n_\text{p} \alpha_x \bigg(\frac{p}{p_{\text{th},x}}\bigg)^\frac{d+1}{2},\\
&P_{\text{L},2}^Z \equiv \alpha_z \bigg(\frac{p}{p_{\text{th},z}}\bigg)^\frac{d_z+1}{2} \leq \alpha_z \bigg(\frac{p}{p_{\text{th},z}}\bigg)^\frac{d(n_\text{p}-n_\text{k})+1}{2}.
\end{split}
\label{eq:cnot_second_part_XX}
\end{equation}
In Eq.~\eqref{eq:cnot_second_part_XX}, the parameters $\alpha_{x,z}$ and $p_{\text{th},x,z}$ are the monolithic fitting parameters introduced in Eq.~\eqref{eq:cnot_first_part_ZZ}.
We define $n_\text{p}$ as the number of surface code patches that fit into the total routing space used, including the two logical patches, $n_\text{k}$ as the number of ``kinks'' (90-degree angles) in the routing path, and $n_\text{s}$ as the number of Bell pair seams used for the path.
Similarly to Eq.~\eqref{eq:cnot_first_part_ZZ}, the factor $d_z/d_x$ describes logical errors that can occur in the long dimension of the rectangular surface code patch. 
Bell pair errors mainly contribute to logical $X$ errors, as the Bell pair seams stretch in the $X$ direction of the patch. 
The parameter $p_\text{Bell}$ represents physical error probabilities as introduced by Bell pairs on the seams. 
The parameters $p_\text{Bell}^*$, $p^*$, and $\alpha_{1,2,3,\text{c}}$ follow from fitting the logical $X$ error probability of a surface code patch with a Bell pair seam along the $X$ direction with our physical error model of Appendix~\ref{app:noise_model}. 
Logical $Z$ errors are approximated by the monolithic model, following Ref.~\cite{jacinto2026network}. 
If the control patch and target patch are part of the same node---i.e., $n_\text{s}=0$---we use the monolithic model for the logical $X$ errors as well. 
The distance $d_z$ of the $\text{M}_{XX}$ surface code patch depends on the exact shape of the routing path. In Eq.~\eqref{eq:cnot_second_part_XX}, we use $n_\text{p}-n_\text{k} \leq d_z/d_x \leq n_\text{p}$ to pessimistically estimate the effect of $d_z$ on the logical error probabilities, where the inequality for $P_{\text{L},2}^Z$ specifically holds for $p \leq p_{\text{th},z}$. 

In our success-probability calculations, we neglect the effects of the conditional logical $X$ and $Z$ corrections, as well as the final $Z$ measurement of the ancillary patch in Fig.~\ref{fig:cnot_figure}. 
Conditional logical Pauli corrections can be implemented via Pauli-frame updates and therefore do not require physical gate operations. 
The final logical $Z$ measurement of the ancillary patch can be implemented destructively by measuring all data qubits of the patch and classically decoding the logical observable. 
In contrast, the joint $ZZ$ and $XX$ measurements require lattice surgery (i.e. $\geq d$ rounds of stabiliser measurements) before decoding the error syndrome, and therefore dominate the logical failure probability. 

The total probability of the logical CNOT to succeed is now given by
\begin{equation}
\bar{\mathbb{P}}_\text{CNOT} \approx (1-\mathbb{P}_{ZZ,\text{rp}})^{s_1} (1-\mathbb{P}_{XX,\text{rp}})^{s_2}. 
\label{eq:success_probability_CNOT}
\end{equation}
In Eq.~\eqref{eq:success_probability_CNOT}, $s_{1,2}$ describes the number of stabiliser measurement rounds used for the $\text{M}_{ZZ,XX}$ operation, respectively.
For rectangular surface code patches or lattice surgery between remote patches, increasing the number of stabiliser cycles beyond the shortest distance $d$ prevents a higher logical error probability caused by more measurement-induced error strings fitting in the time dimension along the prolonged spatial dimension. 
In the models of Eqs.~\eqref{eq:cnot_first_part_ZZ} and~\eqref{eq:cnot_second_part_XX}, a similar effect in the two spatial dimensions is modelled with linear factors $d_x/d_z$ and $d_z/d_x$, respectively, following Ref.~\cite{jacinto2026network}. 
However, literature suggests that the optimal number of stabiliser measurements for lattice surgery between remote patches only scales (at most) logarithmically with the separation width between patches~\cite{domokos2024characterization,litinski2017braiding}. 
For the $\text{M}_{XX}$ operation, we therefore use
\begin{equation}
s_2 = d+\ln\big((n_\text{p}-2)d\big)
\label{eq:cnot_second_part_XX_number_cycles}
\end{equation}
stabiliser measurement cycles. 
In Eq.~\eqref{eq:cnot_second_part_XX_number_cycles}, $n_\text{p}-2$ represents the number of surface code patches that fit in the routing space between the two logical patches. 
For the $\text{M}_{ZZ}$ operation, we simply use $s_1=d$ stabiliser measurement cycles, since we ensured that the two involved patches are always right next to each other. 
The total number of stabiliser cycles of the full CNOT $s_\text{CNOT}=2d+\ln((n_\text{p}-2)d)$ thus only depends on the code distance and the length of the route used to connect the two involved logical patches. 

In our calculations, we assume that multiple non-overlapping CNOT operations $N_\text{CNOT}$ take place in parallel. 
To calculate the probability that all parallel CNOTs succeed, we also include the error probability of idling logical patches. 
Per stabiliser measurement cycle, the probability that one idling surface code logical qubit produces a logical error is given by
\begin{equation}
\mathbb{P}_{1\text{lq}} \approx \alpha_x \bigg(\frac{p}{p_{\text{th},x}}\bigg)^\frac{d_x+1}{2} + \alpha_z \bigg(\frac{p}{p_{\text{th},z}}\bigg)^\frac{d_z+1}{2}.
\label{eq:idling_one_logical}
\end{equation}
We denote $N_\text{log}$ as the number of logical surface code patches present, we define $s_{\text{CNOT},i}$ as the number of stabiliser cycles required to execute an individual CNOT labelled with $i$ and $s_\text{CNOT,tot} \equiv \max_{i=1}^{N_\text{CNOT}} s_{\text{CNOT},i}$ as the total number of stabiliser cycles required to execute all parallel CNOTs. 
The logical error probability $\mathbb{P}_\text{tot}$ of all parallel CNOTs and idling logical patches is now given by
\begin{equation}
\begin{split}
&\mathbb{P}_\text{tot} \approx 1 - \bigg(\prod_{i=1}^{N_\text{CNOT}} \bar{\mathbb{P}}_{\text{CNOT},i}\bigg)\bar{\mathbb{P}}_\text{idling},\\
&\bar{\mathbb{P}}_{\text{CNOT},i} \equiv (1 - \mathbb{P}_{ZZ,\text{rp}})^d (1 - \mathbb{P}_{XX,\text{rp},i})^{s_{\text{CNOT},i}-d},\\
&\bar{\mathbb{P}}_\text{idling} \equiv (1 - \mathbb{P}_{1\text{lq}})^{s_\text{CNOT,tot}(N_\text{log}-N_\text{CNOT})+\sum_{i=1}^{N_\text{CNOT}}(s_\text{CNOT,tot} - s_{\text{CNOT},i})}.
\end{split}
\label{eq:total_success_probability_CNOTs}
\end{equation}
In Eq.~\eqref{eq:total_success_probability_CNOTs}, we have used that the operation $\text{M}_{ZZ}$ is the same for every logical CNOT, but $\text{M}_{XX}$ and $\mathbb{P}_{XX,\text{rp}}$ depend on the length and shape of the patch. 
The last term in the exponent of $\bar{\mathbb{P}}_\text{idling}$ takes into account idling errors for CNOTs that finish before the last CNOT. 
In the first part of the exponent of $\bar{\mathbb{P}}_\text{idling}$, we use $N_\text{log}-N_\text{CNOT}$ and not $N_\text{log}-2N_\text{CNOT}$ because there is always one logical patch idling during each CNOT: first the target patch and then the control patch.

\section{Fixed $t_{\mathrm{Bell}}$: varying Bell state fidelity}\label{sec:fixed_duration}

\begin{figure*}[t!]
\centering
\centering
\begin{minipage}{\textwidth}
\includegraphics[width=\textwidth]{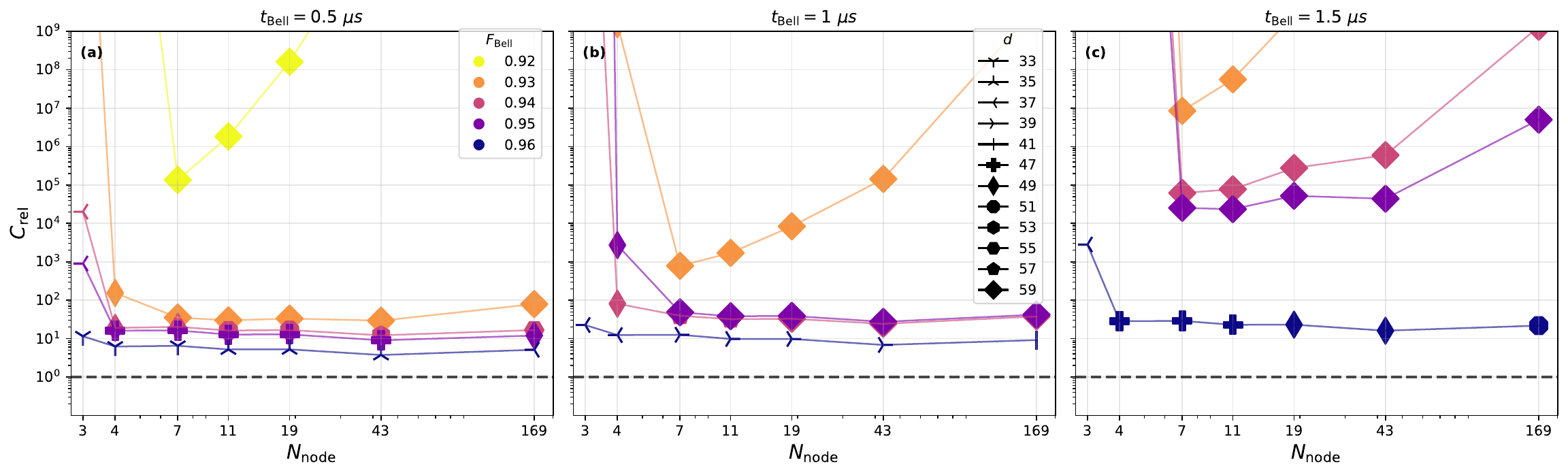}
\caption{\textbf{Fixed $t_{\mathrm{Bell}}$ with varying fidelity.} Relative space-time costs are shown for varying Bell state parameters with respect to the monolithic baseline value (indicated by the black striped line). Yellow to purple colouring indicates a lower to higher fidelity. The $y$-axis is truncated for readability; omitted datapoints lie above the plotting limit of $C = 10^9$. Fidelities $97\%$, $98\%$ and $99\%$ are not shown because of the overlap with the results for $96\%$.}
\label{fig:fixed_duration}
\end{minipage}
\end{figure*}

We also consider the impact of Bell-state fidelity on the relative space-time cost ($C_{\mathrm{rel}} = C_{\mathrm{dist}}/C_{\mathrm{mono}}$) for fixed Bell-pair generation times ($t_{\mathrm{Bell}}$), as shown in Fig. \ref{fig:fixed_duration}. Across all evaluated regimes, the distributed approach carries a higher space-time cost than the effectively monolithic baseline, revealing an inherent networking overhead. However, the severity of this overhead is dictated by specific performance thresholds. For fast links where entanglement generation outpaces the QEC cycle time (Fig. \ref{fig:fixed_duration}a), fidelity is the primary bottleneck. We observe a sharp transition at $F_{\mathrm{Bell}} \approx 92\%$; below this threshold, seam-induced logical errors compound, driving the space-time cost up by many orders of magnitude as node count increases. Above \(93\%\), the interconnect is no longer the dominant bottleneck, and the improved seam fidelity allows for smaller code distances. As interconnect latency approaches and exceeds the QEC cycle time, shown in Fig. \ref{fig:fixed_duration}b for $t_\textrm{Bell} = 1\mu s$ and Fig. \ref{fig:fixed_duration}c for $1.5~\mu s$, idling errors accumulate during each cycle. This shifts the viability threshold toward progressively higher fidelities to prevent superexponential cost scaling. 

Our results show that expanding the network size can mitigate suboptimal interconnects by reducing routing congestion and enhancing parallelism. In small networks (Fig. \ref{fig:fixed_duration}b), limited routing paths bottleneck the entanglement distribution, and adding nodes resolves this for intermediate-to-high fidelities ($\ge 94\%$). However, if the interconnect fidelity is too low (e.g., the \(93\%\) curve in Fig. \ref{fig:fixed_duration}b), the requisite increase in inter-node seam traversals overwhelms any routing benefits, penalising larger networks.

\section{Near-ideal seams}\label{app:30ns}

\begin{figure}[t]
\centering
\includegraphics[width=\columnwidth]{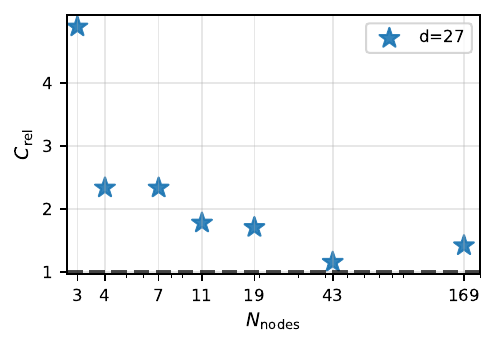}
\caption{\textbf{Near-ideal Bell links.} Relative expected qubit$\times$time cost \(C\) versus node count $N_{node}$ for \(t_{\mathrm{Bell}}=30\)~ns and Bell state fidelity \(99.95\%\), comparable to local two-qubit gates. The monolithic baseline is again shown using a dotted black line. Even in this limit, distributed execution does not collapse to monolithic performance. Note that this graph has no logarithmic axes.}
\label{fig:30ns_local_like}
\end{figure}

In Fig.~\ref{fig:30ns_local_like}, we set the Bell link fidelity and rate equal to the fidelity and time a local two-qubit gate specifications, i.e. (\(t_{\mathrm{Bell}}=30\)~ns, \(99.95\%\) fidelity).
This decouples the connectivity-induced overhead from overhead induced by chosen Bell pair specifications, allowing us to see the sole effect of limited connectivity on cost performance with compiled long-range lattice surgery operations.

In this limit, $C_\mathrm{rel}$ decreases with increasing node count across most of the sweep, reflecting that routing freedom and parallelisation improve as seam errors no longer dominate and idling during Bell-pair creation becomes negligible. 
$C_\mathrm{rel}$ approaches 1 in some cases, showing practically equal performance between the distributed and effectively monolithic set-up.
As the number of nodes increases, additional routing opportunities for qubits on separate nodes become available, forming a more complete topology (which is more akin to the effectively monolithic set-up).

However, the distributed architecture still does not precisely equal or trump the effectively monolithic baseline, even with equal interconnect quality. 
Networking via entanglement distribution with Bell pairs introduces errors that do not vanish under such parameter improvements as derived in the distributed lattice-surgery error model presented in App.~\ref{app:cnot}.
Pushing $N_{node}$ to very large values does not further improve performance because, as the partition becomes finer, all two-qubit gate interactions require the use of the (several) seams, which add idling time ($30\,\mathrm{ns}$) in every QEC cycle for every seam-mediated lattice surgery, and further strain the decoder.

\end{document}